\documentclass[12pt, a4paper]{article}
\usepackage[font=footnotesize]{caption}
\usepackage[utf8]{inputenc}
\usepackage{graphicx}
\usepackage{authblk}
\usepackage{booktabs}
\usepackage{verbatim}
\usepackage{url}
\usepackage{caption}
\usepackage{subcaption}
\usepackage{indentfirst}
\usepackage{url}
\usepackage{amssymb}
\usepackage{amsmath}
\title{Hierarchical Cutting of Complex Networks Performed by Random Walks}

\author{Alexandre Benatti$^1$ and Luciano da F. Costa$^2$}

\affil{
$^1$Institute of Mathematics and Statistics - DCC \\
University of S\~ao Paulo \\
Rua do Mat\~ao, 1010, S\~ao Paulo, SP 05508-090 Brazil 
\\ \vspace{0.5cm}
$^2$S\~ao Carlos Institute of Physics - DFCM \\
University of S\~ao Paulo \\
Av.~Trabalhador S\~ao-Carlense, 400, S\~ao Carlos, SP 13566-590 Brazil
}

\date{15th Feb. 2024}

\begin{document}

\maketitle

\begin{abstract}
Several interesting approaches have been reported in the literature on complex networks, random walks, and hierarchy of graphs. While many of these works perform random walks on stable, fixed networks, in the present work we address the situation in which the connections traversed by each step of a uniformly random walks are progressively removed, yielding a successively less interconnected structure that may break into two components, therefore establishing a respective hierarchy. The sizes of each of these pairs of sliced networks, as well as the permanence of each connected component, are studied in the present work. Several interesting results are reported, including the tendency of geometrical networks sometimes to be broken into two components with comparable large sizes.
\end{abstract}

\section{Introduction}\label{sec:introduction}

The areas of complex networks (e.g.~\cite{newman2018networks, barabasi2013network,costa2007characterization,costa2011analyzing,boccaletti2006complex}) and random walks (e.g.~\cite{barber1970random,lovasz1993random,spitzer2013principles,xia2019random}), which are interesting on themselves, have often been brought together (e.g.~\cite{noh2004random,baronchelli2008random,zhang2013random,carletti2020random}) in several interesting approaches reported in the literature. Typically, random walks are performed on given fixed networks as a means of: (a) studying the interplay between the obtained dynamics and the topology of networks (e.g.~\cite{da2007correlations}); (b) exploring networks (e.g.~\cite{da2007exploring,starnini2012random}); (c) analyzing and modeling synthetic and/or real-world complex systems (e.g.~\cite{masuda2017random,luo2006application}); and (d) inferring properties (including modularity) of given networks (e.g.~\cite{pons2005computing,lambiotte2014random,rosvall2009map}). Other approaches have also used random walks for additional purposes, including to generate complex networks (e.g.~\cite{costa2007knitted}), to compare networks (e.g.~\cite{lu2014complex}), or as a subsidy for characterizing the resilience of networks to attacks (e.g.~\cite{turchetto2023random}). Networks have also been performed on time-varying networks (e.g.~\cite{perra2012random,starnini2012random}).

In the present work, we focus attention on the interesting situation in which, given a specific complex network, a random walk is respectively performed so that all traversed connections (edges) are removed from the network as a consequence of the agent displacements. As a consequence, the nodes of the original network become less and less interconnected, up to a point where the network breaks into two or more connected components. Slicing graphs, known as \emph{graph cutting}, corresponds to an interesting subject studied in graph theory (e.g.~\cite{gross2018graph}), though these studies tend to focus on static graphs.

Several aspects of interest are implied by the above mentioned issue, including the characterization of the changes of the topological properties undergone as the networks are gradually sliced, with a special interest in the partitioning of the network into separated connected components respectively to different types of networks and random walks. Interestingly, as the network is progressively disassembled (or dismantled) into disconnected groups (connected components), a respective \emph{hierarchy} is established where each current component corresponds to a node while the respectively disconnected portions are represented as respective leaves, therefore establishing a hierarchical branch. The hierarchical aspects of graphs and networks have also received attention from the literature (e.g.~\cite{sales2007extracting,clauset2008hierarchical,benatti2023recovering,barabasi2003hierarchical,corominas2013origins}) as a means of better understanding their respective topology and properties.

To address the above mentioned issues provides the main objective of the present work. In addition to the several theoretical aspects underlying this investigation, the addressed issue of gradual cutting of complex networks performed along random walks is related to a large number of practical and real-world situations. Of particular practical interest is the study of the sizes of the pairs of disconnected components obtained along the slicing implemented by the random walk. Two main situations can occur: (a) one of the components in each pair is always very small (near one node), while the other component is larger, which is henceforth called \emph{sequential dismantling}; and (b) some pairs of disconnected components have comparable (and possibly large) sizes, which is in this work called \emph{abrupt dismantling}. 

The distinction between the two types of network dismantling identified above is of particular interest in cases where a single agent is employed in the random walk, therefore implying only one of each of the pairs of possible disconnected components to be taken along the random walk. In the case of \emph{sequential dismantling}, the whole network will be progressively disassembled into successively smaller instances, without leaving behind substantial lumps of the structure. In the case of \emph{abrupt dismantling}, considerable portions of the original network can be prevented from being disassembled because the random walk agent can only proceed into one of two disconnected components of comparable large sizes.

It follows from the above discussion that networks that, among other implications, the exploration or usage of the network performed by a respective sequential dismantling will be more effective, as no substantial portion of the structure will be overlooked.

Another interesting aspect related to the above type of gradual cutting of networks concerns the characterization of how long each involved connected component lasts until being broken into two respective connected components. This aspect of component permanence is also studied here.

This work starts by presenting basic concepts related to complex networks and discrete random walks. Then, the approach to hierarchically cutting networks is described, which is followed by experimental results involving three types of networks and respective discussions. Several interesting results are presented, including the tendency of geometrical networks to be sometimes broken into two relatively large connected components, which is substantially less likely to occur in the uniform and preferentially interconnected types of networks.

\section{Basic Concepts}\label{sec:methods}

Complex networks are composed of \emph{nodes} (or \emph{vertices}) that are connected by \emph{edges} (or \emph{links}), which are often used to represent interactions among the nodes. The \emph{adjacency matrix} ($A$) is a typical representation of a network. Each entry in this matrix indicates whether there is an edge between two respective nodes. If node $i$ is connected to node $j$, the corresponding entry ($A_{ij}$) in the adjacency matrix is set to one, otherwise, it is set to zero.

The \emph{size} of a given complex network is henceforth understood as corresponding to its number of nodes. 

The \emph{degree} of a node is determined by the number of edges attached to it. The degree ($k_i$) of node $i$ can be expressed using the adjacency matrix.

\begin{equation}
    k_i = \sum_{j=1}^N A_{ij},
\end{equation}
where $N$ represents the total number of nodes, and $A_{ij}$ represents the element $(i,j)$ in matrix $A$.

There are various models of complex networks, each characterized by specific topology, symmetry, and connectivity. Features characterizing these properties can help in understanding the network topology and dynamics. In this work, we considered three different network models: Erdős–Rényi (ER)~\cite{erdos1959random}, Barabási–Albert (BA)~\cite{barabasi1999BA}, and a Geometric Graph (GEO) -- e.g.~\cite{riedinger1988delaunay}.

The ER model generates random graph connections with uniform probability. More specifically, nodes are connected randomly with a fixed probability. ER graphs exhibit a binomial degree distribution (e.g.~\cite{newman2001random}), which implies many of the degrees to be similar to the respective average.

The BA model employs a preferential attachment mechanism, whereby nodes with higher degrees are more likely to receive new links. As the network expands, new nodes attach to existing nodes based on their current connectivity. BA graphs are characterized by a power-law degree distribution, indicating that a few nodes have significantly more connections than others.

Geometrical complex networks are characterized by nodes occupying specific geometric positions so that the proximity and/or adjacency between nodes can influence the respective interconnections. In the present work, we consider a specific geometric network obtained by considering the Voronoi diagram obtained from the nodes position (e.g.~\cite{riedinger1988delaunay}). More specifically, here the nodes are distributed uniformly within a two-dimensional lattice incorporating a small uniformly distributed spatial displacement~\cite{benatti2023simple}, the Voronoi diagram is obtained, and every pair of nodes that results adjacent in the Voronoi tesselation is connected.

Given a complex network, any subset $S$ of nodes is said to be a \emph{connected component} provided: (a) any of these nodes can be accessed by a path initiating at a node in that set; and (b) all the network nodes not included in $S$ are unreachable from that set.

Though all complex networks considered in the present work are non-directed, the described concepts and methods can be readily extended to directed networks.

Given a network, time-discrete \emph{random walks} can be respectively performed by one or more abstract agents moving along connected nodes. In the present work, in the case of a single agent, we adopt a uniform random walk in which, after starting at a specific node, the moving agent proceeds to neighboring nodes with uniform probability. Agents can make a move (i.e.~traverse an edge) at each discrete time step. Random walks can also be performed on weighted networks, e.g.~by considering normalized transition probabilities proportional to the weights.

\section{Methodology}\label{sec:Methodology}

In this work, we consider random walks where each link traversed by a moving agent is deleted. Given that this dynamics can lead to the current network being disconnected into a pair of components, a new moving agent is assigned so that each of the two components continues to be traversed by a respective agent, and so on until the network is completely disassembled. The successive breaking of the network establishes a respective hierarchy that can be represented as a binary tree, namely a tree where all branches have one or two leaves. Observe that the adopted specific procedure leads only to a component being broken into two connected portions, and not 3 or more new components.

\begin{figure}
  \centering
     \includegraphics[width=.7 \textwidth]{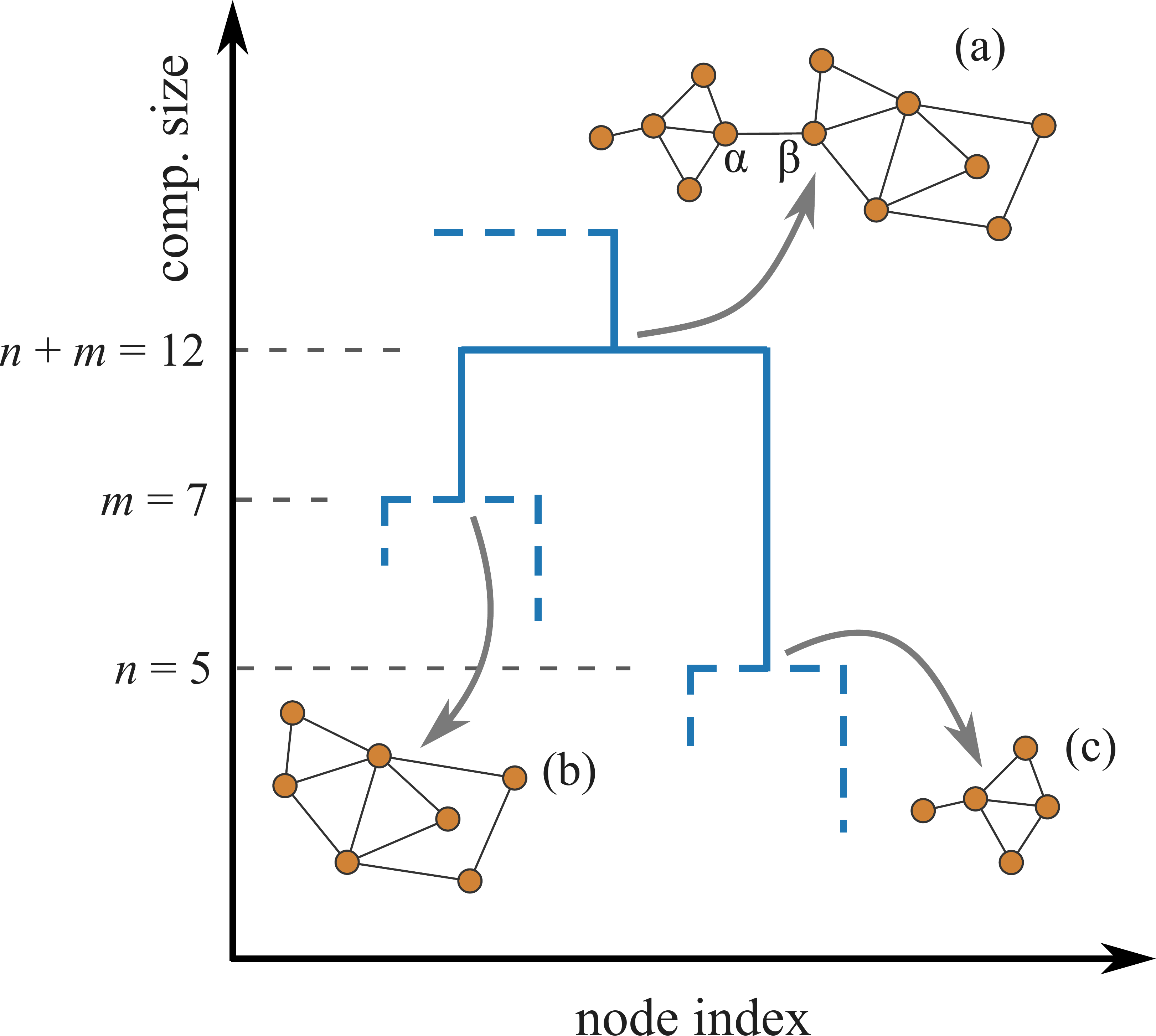}
   \caption{Illustration of the basic event in the considered gradual slicing of a network by a respective random walk. One of the obtained connected components (a) is broken into two connected components (b) and (c) by an agent moving from node $\alpha$ to node $\beta$.}\label{fig:branch}
\end{figure}

Figure~\ref{fig:branch} illustrates the basic branching event involved in the experiments described in this work. It shows one instance of the network (a) being broken into two respective connected components (b) and (c) as a consequence of an agent proceeding from node $\alpha$ to $\beta$ (or from $\beta$ to $\alpha$). This event is represented by the respective branch shown in solid blue lines. Observe that the size $n+m$ of the component (a) corresponds to the height of the parent branching. The heights of components (b) and (c) define the lower extremity of the child branches, which are respectively $m=7$ and $n=5$ in this example. A branch characterized by having child branches with similar lengths is henceforth said to be \emph{balanced}. In case child branches with comparable lengths are also short, the two components will not only be balanced but also have large sizes.

The whole branching structure obtained from the original network up to its complete decomposition into one-node components corresponds to a binary tree which is henceforth understood as a \emph{dendrogram}.

It is important to observe that, given the random nature of the edge removals implemented as described above, several distinct dendrograms can be obtained from the same original complex networks.

\begin{figure}
  \centering
     \includegraphics[width=.99 \textwidth]{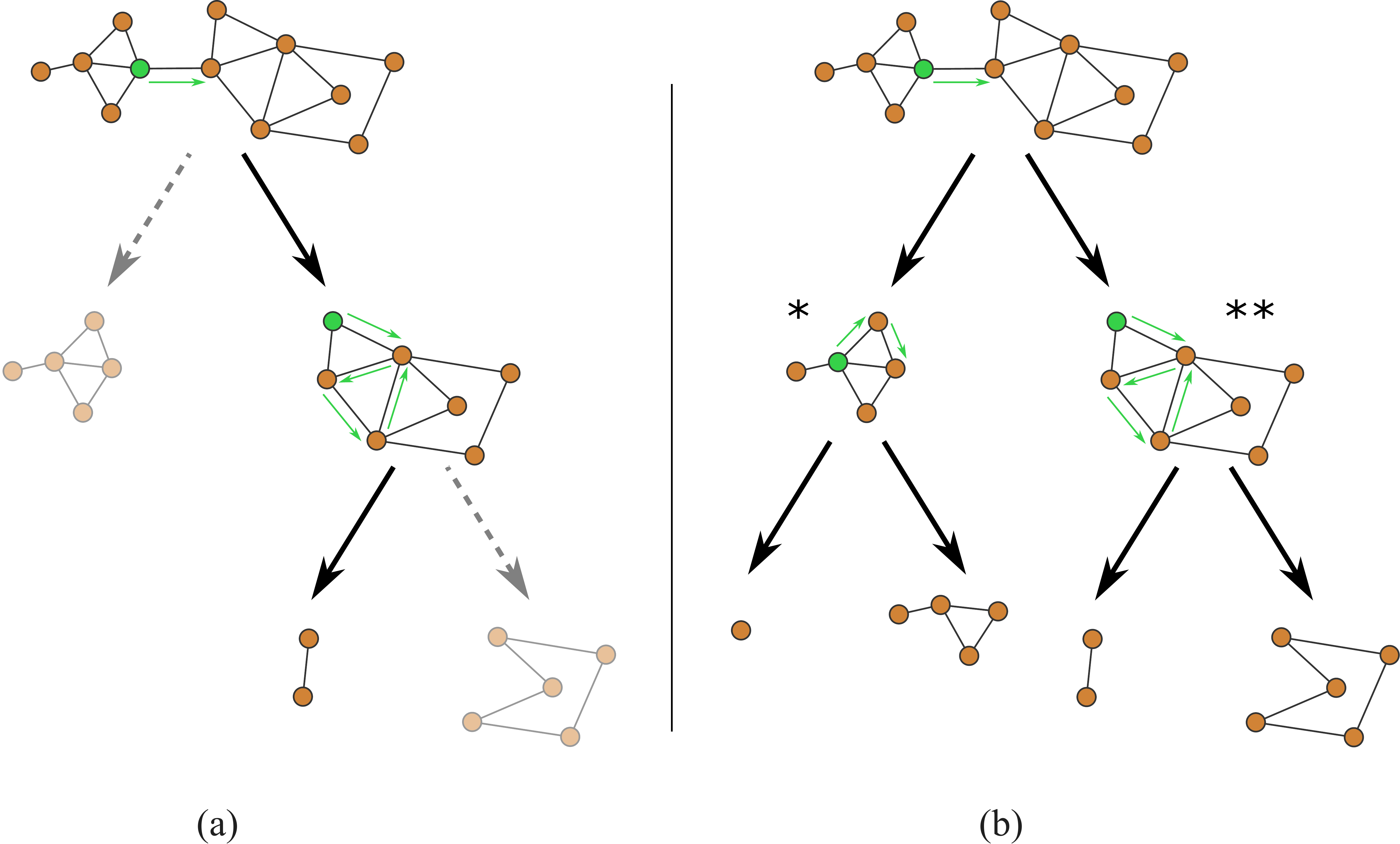}
   \caption{The two types of cutting dynamics considered in the present work when a current component is broken into two new connected components. In the \emph{sequential walk} (a), a single agent remains in only the new connected component into which it moves but is assigned to a randomly chosen node. \emph{Parallel walk} (b) involves assigning agents to randomly chosen nodes of each of the two new connected components. The green arrows show the agent movements within the connected components up to their respective separation.}\label{fig:single_mult}
\end{figure}

Though the dendrogram in Figure~\ref{fig:branch} is intrinsically `parallel', in the sense of several possible branchings being included, it is possible to traverse the respective dendrogram in two main ways considered in the present work: (a) \emph{sequential walk}, illustrated in Figure~\ref{fig:single_mult}(a), in which a single agent remains in only the child branch to which it moves into; and (b) \emph{parallel walk}, shown in Figure~\ref{fig:single_mult}(b), which assigns an agent to each of the two new connected components. In both cases, the agent is assigned into a node chosen uniformly among the nodes in the respective connected component. As a consequence, only one of the paths is obtained in the sequential approach, extending from the root of the dendrogram to one of its termination leaves. Several branches are otherwise covered by the parallel approach.

In addition to considering the balance between the obtained pairs of connected components, it is also of particular interest to quantify the number of steps along the walk performed by each agent from it being assigned to a component up to the breaking of the latter into two new components. This measurement is respectively called \emph{permanence} $P$. As an example, the connected components labeled as (*) and (**) in Figure~\ref{fig:single_mult} have respective permanences $P=2$ and $4$ time steps.

Given a particular network and dynamics, larger values of $P$ will indicate that the respective components remain connected along a greater period (steps), being, therefore, more resilient (in this aspect) to the progressive slicing performed by the moving agent(s). 

\begin{figure}
  \centering
     \includegraphics[width=.6 \textwidth]{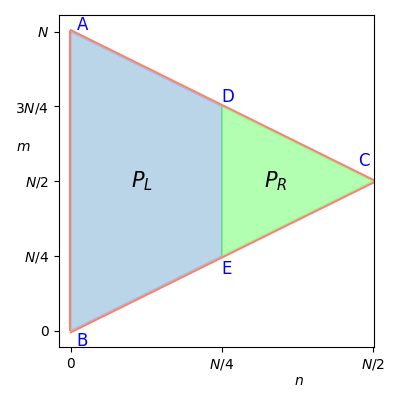}
   \caption{The possible pairs $(n,m)$ that can be obtained by the adopted network slicing procedure are restricted to the triangular region $ABC$. Henceforth, we divide this region into two respective sub-regions ADEB (blue) and DCE (green), with the latter region corresponding to pairs of broken components having more comparable sizes. Observe that $N$ is the number of nodes in the original network.}\label{fig:Geometry}
\end{figure}

Given that we are particularly interested in studying the size of the obtained pairs of connected components, more specifically in identifying pairs with comparable sizes, we consider the two-dimensional discrete space defined by representing the sizes $n$ and $m$, with $1 \leq n \leq m$, of each pair of components into the horizontal and vertical axes, respectively. Figure~\ref{fig:Geometry} depicts the diagram of the possible points that can be obtained during the adopted slicing dynamics, which corresponds to the triangular region ABC (filled in blue and green). Observe that this region is delimited by the three following line segments: (i) AB for the reason that $n, m > 0$; (ii) BC because $n \leq m$; and (iii) AC, originating from the constraint $n+m \leq N$.  

Given a specific distribution of component pairs obtained by the described methodology, it becomes possible to estimate the probabilities of having less or more balanced connected pairs, which are indicated as $P_L$ and $P_R$, respectively. These probabilities correspond to the integration of the overall density within the regions $ADEB$ and $DCE$.

\section{Experiments and Discussion}\label{sec:results}

The experiments consisted of performing the network cutting dynamics to three types of complex networks, namely ER, BA, and GEO (specific case of Voronoi). All considered networks have the same number of nodes $N=100$ and similar average degrees $\left< k \right> \approx 5.7$. Similar results have been observed for other values of $N$. In the case of ER networks, the largest connected component is taken as the initial structure to be traversed by the random walks. The average degrees have been assumed as being approximately equal to 5.7. This is the experimentally estimated value of the average degree in the case of the considered Voronoi geographical network generated from a perturbed lattice of 10 $\times$ 10.

The sizes $n$ and $m$ obtained at every cutting, as well as the permanence $P$ of each component, were recorded for subsequent analysis.

The results considered in this work included, for each of the three considered network types: (a) the histogram distribution (average $\pm$ standard deviation) of the duration of sequential random walks; (b) the histogram distribution (average $\pm$ standard deviation) of the networks node degrees; (c) the $m \times n$ scatter plots considering multiple agents; (d) examples of dendrograms obtained by parallel random walks; and (e) histograms of the permanence times considering multiple agents. All these results take into account the results from the above described experiments considering all the networks and starting nodes respectively to each of the three network types.

The experimental results and respective discussions concerning the duration and hierarchy aspects are described in the two subsequent subsections.

\subsection{Sequential Random Walk Duration}

\begin{figure}
  \centering
     \includegraphics[width=.32 \textwidth]{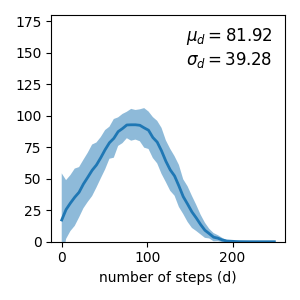}
     \includegraphics[width=.32 \textwidth]{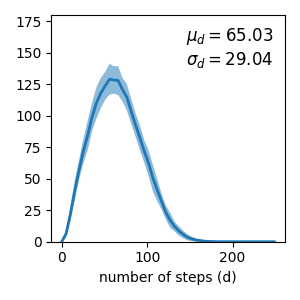}
     \includegraphics[width=.32 \textwidth]{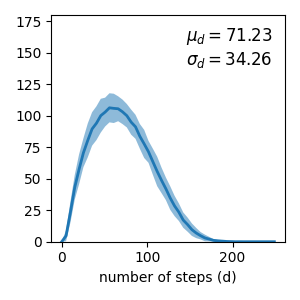}\\
     (a) \hspace{3.7cm} (b) \hspace{3.7cm} (c)
   \caption{The average $\pm$ standard deviation of the total duration of each sequential random walk (single agent) respectively to: (a) ER network, (b) BA network, and (c) GEO network. These results consider 10,000 walks starting at different nodes (randomly chosen) for 50 networks of each type.}\label{fig:hist}
\end{figure}

Figure~\ref{fig:hist} presents the average $\pm$ standard deviation of the total duration of sequential random walks performed by single agents on~\ref{fig:hist}(a) ER,~\ref{fig:hist}(b) BA, and~\ref{fig:hist}(c) GEO types.  These duration values correspond to the number of steps from the beginning of each random walk (single agent) up to its respective completion.

The smallest duration values resulted in the case of BA networks, also presenting the narrowest distribution. This can be possibly accounted for by the fact that a hub is soon reached by a random walk, with the agent subsequently moving into a separated portion of the network from which it becomes unlikely or even impossible to return to the hub, therefore precluding the visit to the several other nodes, which tend to be attached to the hub in BA networks. 

\begin{figure}
  \centering
     \includegraphics[width=.325 \textwidth]{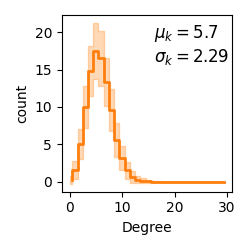}
     \includegraphics[width=.325 \textwidth]{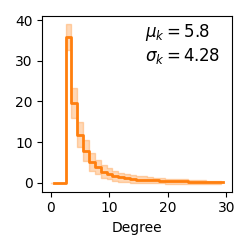}
     \includegraphics[width=.325 \textwidth]{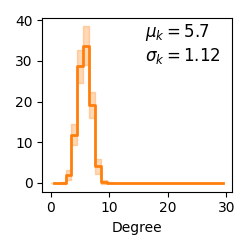}\\
     (a) \hspace{3.7cm} (b) \hspace{3.7cm} (c)
   \caption{Degree distribution (average $\pm$ standard deviation, considering 500 networks of each type) for: (a) ER network, (b) BA network, and (c) GEO network.}\label{fig:deg}
\end{figure}

As could be expected, similar duration values have been observed for the ER and GEO types of networks, which have mostly similar degree distributions. However, the ER networks tended to have more nodes with degrees equal to 1, 2, and 3 which, as illustrated in Figure~\ref{fig:deg}, once visited, tend to shorten the random walks. 

\subsection{Cutting Hierarchy}

In this section, experimental results concerning the balance of the sizes $m$ and $n$ of the broken components ($n \leq m$) along parallel random walks, as well as the permanence of each component are presented and discussed.

\begin{figure}
  \centering
     \includegraphics[width=.325 \textwidth]{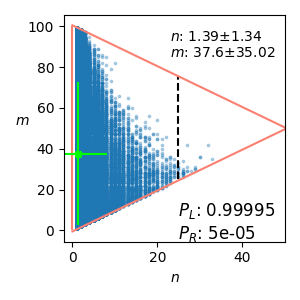}
     \includegraphics[width=.325 \textwidth]{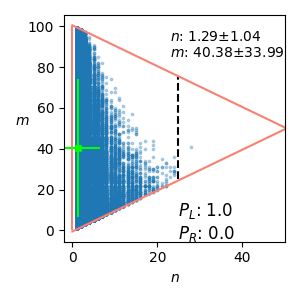}
     \includegraphics[width=.325 \textwidth]{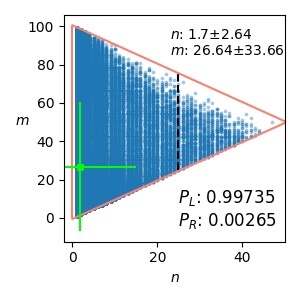}\\
     (a) \hspace{3.7cm} (b) \hspace{3.7cm} (c)
   \caption{The scatterplot diagrams of $n \times m$ obtained for (a) ER network, (b) BA network, and (c) GEO network. The green cross-hair indicates the average and standard deviations (shown magnified by a factor of 5$\times$ along the horizontal axis for the sake of better visualization) of the respectively obtained density. The dashed line separates the two regions respective to the probabilities $P_L$ (left-hand side) and $P_R$ (right-hand side). Interestingly, the distribution extends much further to the right-hand side in the case of the GEO networks (c). A total of 5,000 networks of each type have been considered.}\label{fig:scatter}
\end{figure}

Figure~\ref{fig:scatter} shows the scatterplot diagram (as discussed in Section~\ref{sec:Methodology}) of the tuples $(n,m)$ obtained for the ER (a), BA (b), and GEO (c) complex networks. Also shown are the average $\pm$ standard deviation of the values of $n$ and $m$, as well as the probabilities $P_L$ and $P_R$.

Interestingly, the scatterplot obtained for the GEO networks (c) resulted markedly distinct from those obtained for the ER and BA structures (a,b). More specifically, the distribution of points obtained for the GEO networks extends more widely within the bounding triangle, especially in the smaller triangle on the right-hand side, which corresponds to more balanced component sizes of relatively large sizes. Although the $P_R$ observed for the GEO case is only slightly larger than the probabilities $P_R$ observed for the ER and BA networks. At the same time, $P_R$ is nearly zero in these two latter types of networks. This result turns out to have special importance because it indicates that pairs of connected components of comparably large sizes can be obtained along the random walks in this type of network with a substantially larger probability than in the case of the ER and BA networks. 

\begin{figure}
  \centering
     \includegraphics[width=1. \textwidth]{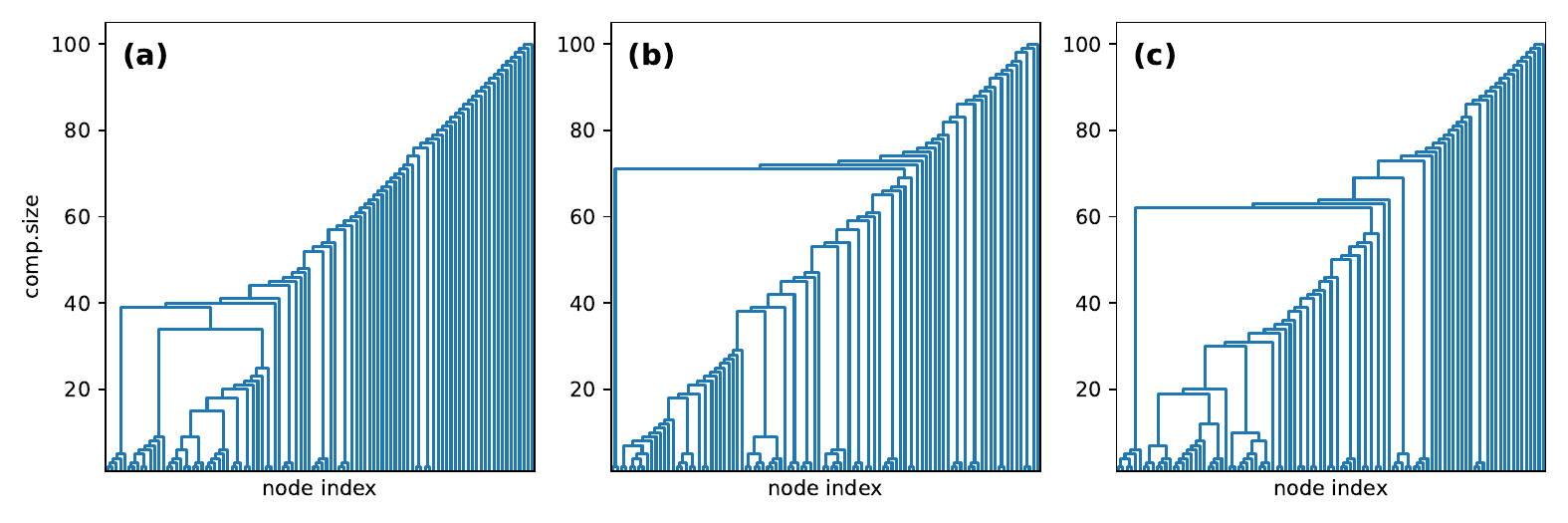}
     \includegraphics[width=1. \textwidth]{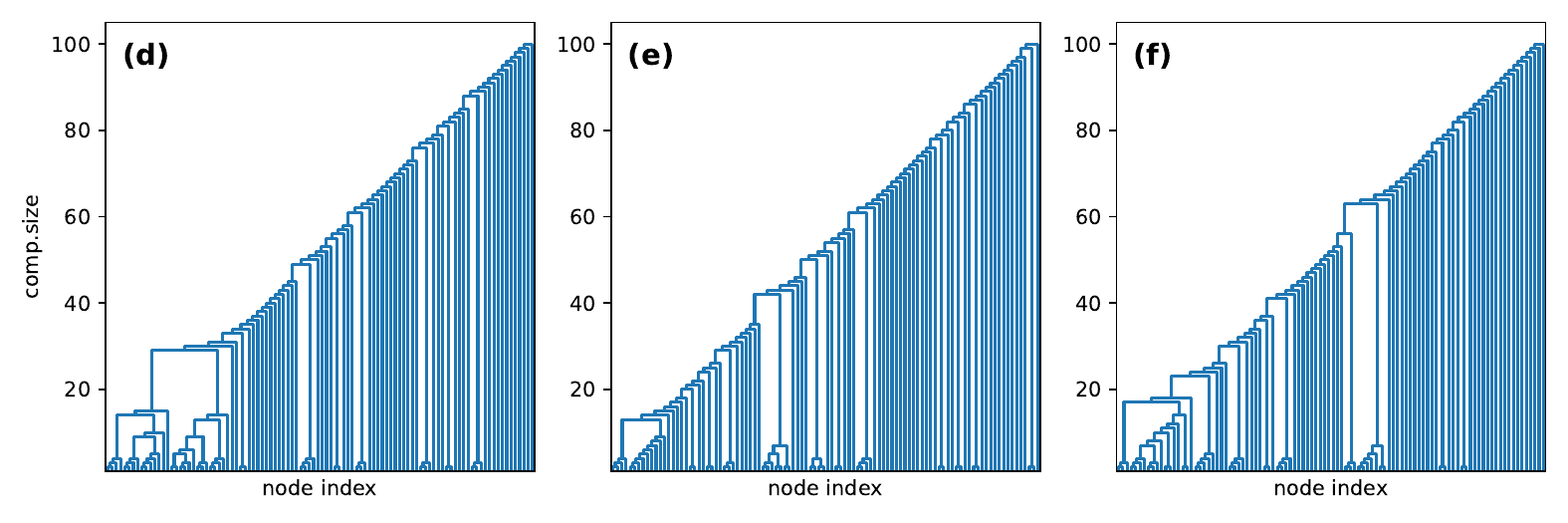}
     \includegraphics[width=1. \textwidth]{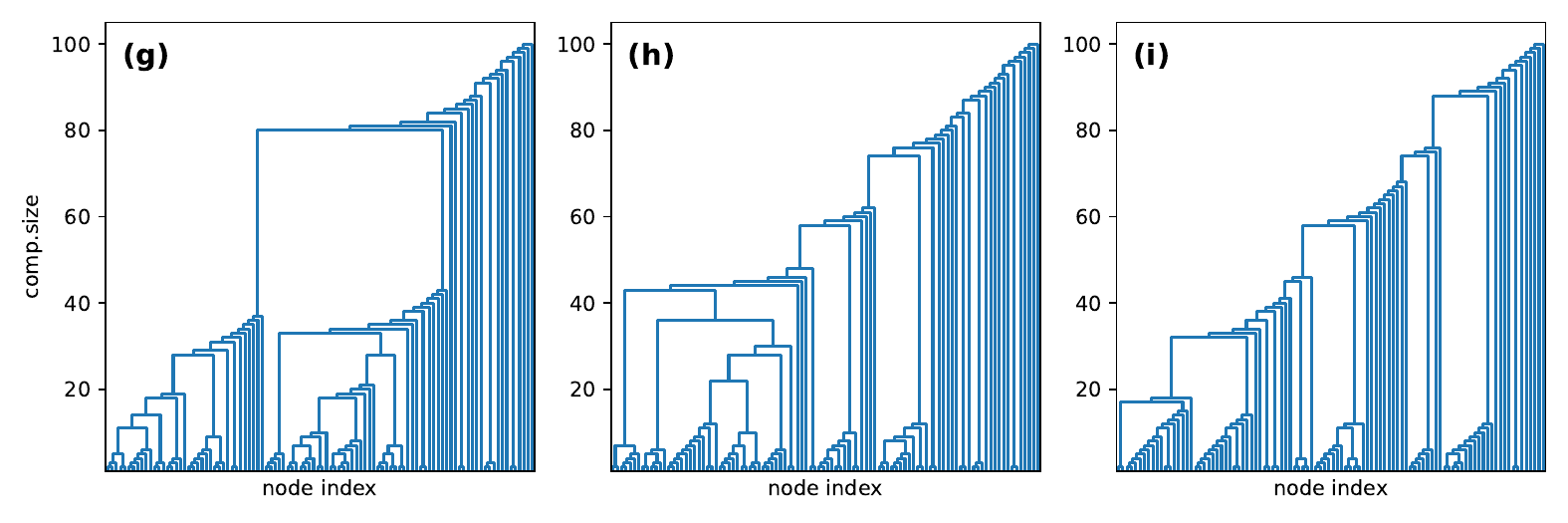}
   \caption{Examples of dendrograms obtained for the ER (a-c), BA (d-f), and GEO (g-i) types of networks. The vertical axis corresponds to the \emph{size} of the connected components during the parallel random walks. Distinct overall structures can be readily observed. The BA networks tend to lead to the most sequential (chained) structures, followed by the ER networks. The GEO structures are characterized by more balanced sizes of pairs of connected components, with substantial branching being observed for relatively large values of component sizes.}\label{fig:dendrogram}
\end{figure}

Examples of dendrograms obtained in the parallel random walks performed on the three considered types of complex networks are illustrated in Figure~\ref{fig:dendrogram}.

The markedly distinct set of connected components appearing in the case of the GEO networks can be readily observed in the respective dendrogram examples. Observe also that some of the branches, in this case, tend not only to be more balanced (similar lengths of child branches) but also relatively larger, involving components with 60 or more nodes. Also of interest is the gradual \emph{chaining} of partitions often observed in the case of the BA networks, in which components containing just one (or a few) nodes are progressively separated from the original network.

A possible explanation for the larger pairs of connected components sometimes appearing in the adopted GEO networks concerns the fact that this type of network is not small-world, while the two other types are. Random walks in non small-world networks tend to be more localized, as there are several nodes that are relatively distant from the moving agent, which is not the case for ER and BA networks. More localized networks have an enhanced probability of acting only on portions of the network providing bridges between other regions, possibly increasing the branching probability.

\begin{figure}
  \centering
     \includegraphics[width=.7 \textwidth]{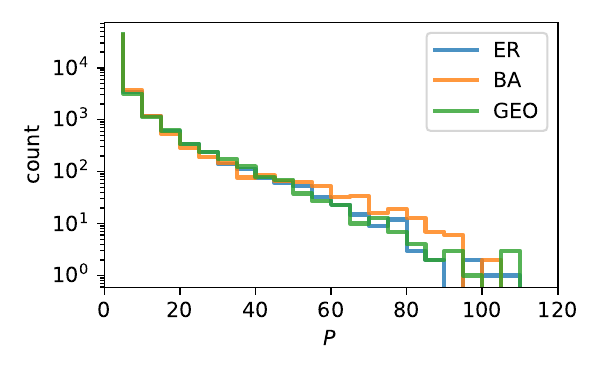}
   \caption{The distributions (histograms) of the permanence $P$ obtained from each of the three network types resulted mostly similar.}\label{fig:permanence}
\end{figure}

Figure~\ref{fig:permanence} depicts the histograms (average $\pm$ standard deviation) of the permanence times $P$ observed for each of the three network types.

These results indicate that similar permanence values are obtained for any of the three considered complex network types and configurations. Thus, interestingly, though larger pairs of components can be more frequently observed in the considered GEO networks, in general, the components of any of the considered networks tend to last for about the same time.

\begin{figure}
  \centering
    \includegraphics[width=1. \textwidth]{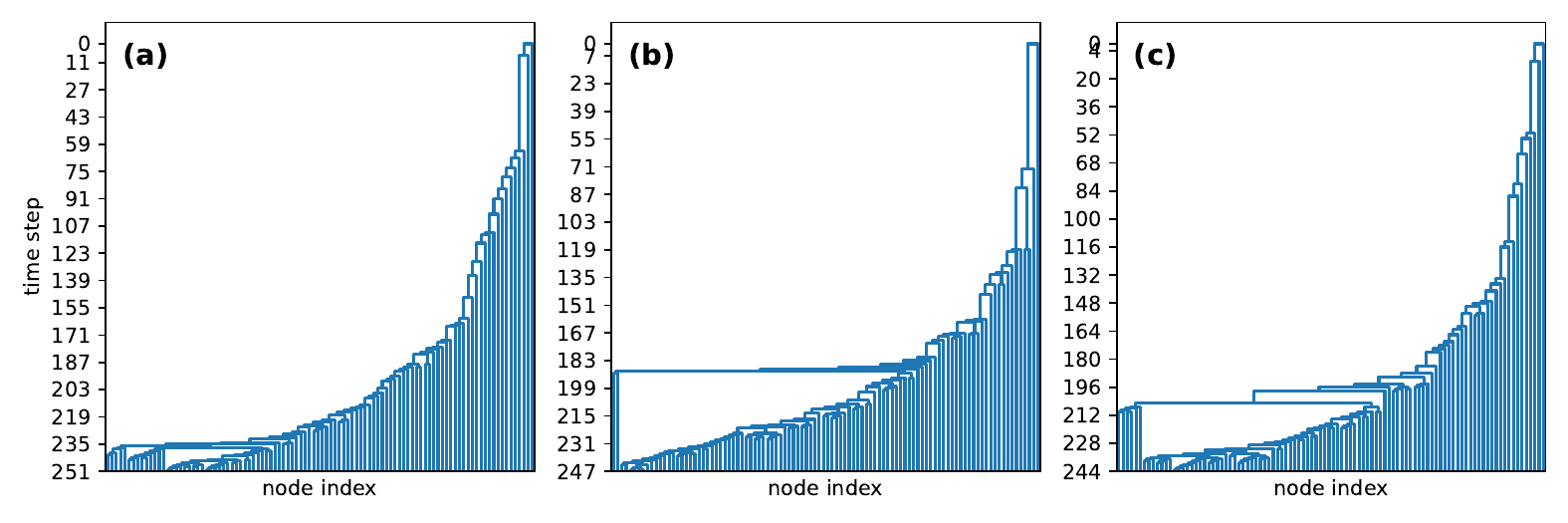}
    \includegraphics[width=1. \textwidth]{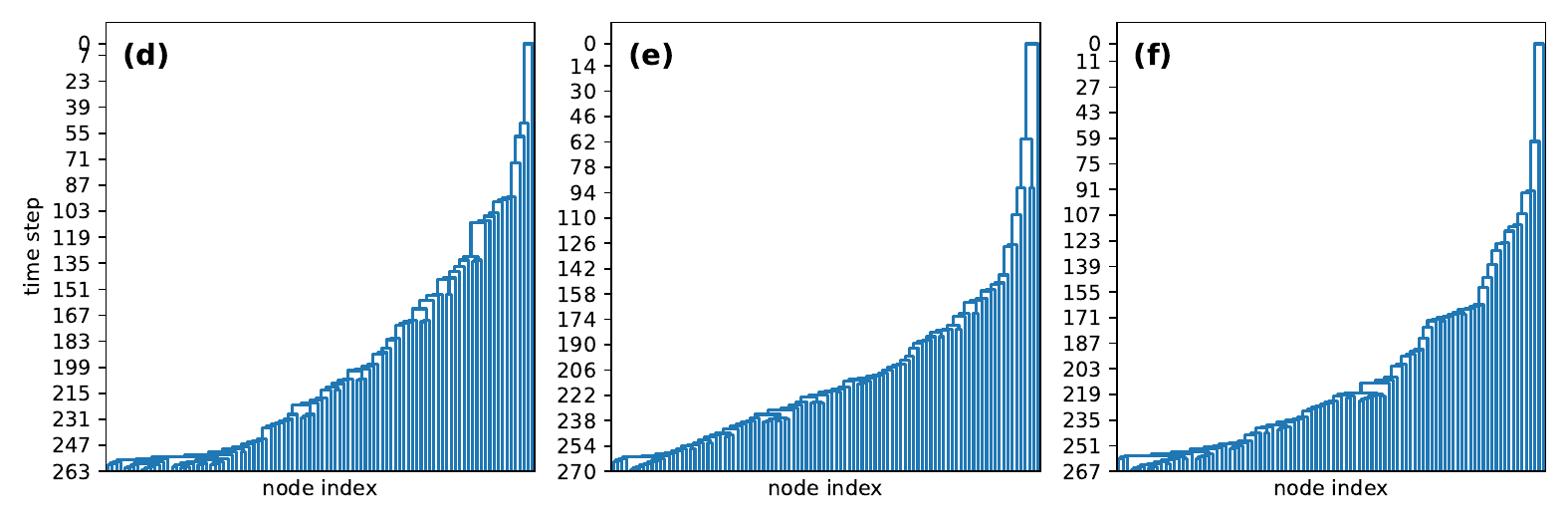}
    \includegraphics[width=1. \textwidth]{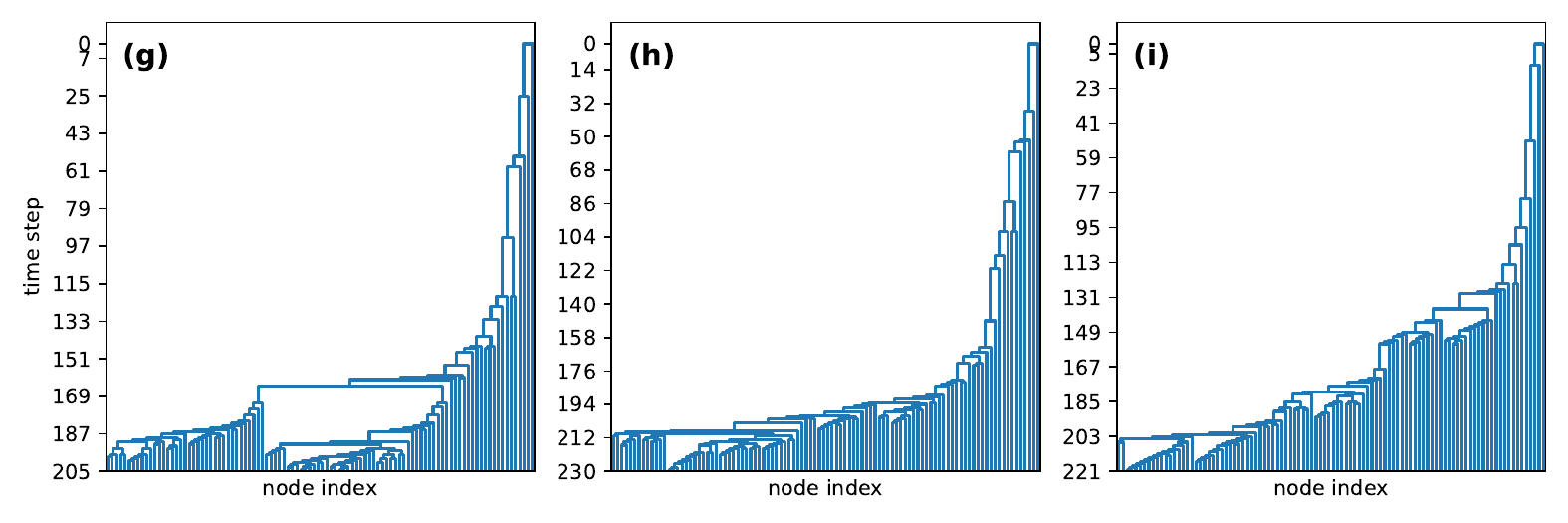}
   \caption{Dendrograms respective to the previous experiments shown in Fig.~\ref{fig:dendrogram}, but now having the vertical axes to correspond to the time steps of the discrete-time random walks instead of the component sizes. A tendency of the larger components in the GEO model to have larger permanence can be observed.}\label{fig:dendrogram_perm}
\end{figure}

Another aspect of particular interest regarding the hierarchical cuttings experimental results regards the permanence of the respectively obtained components. Figure~\ref{fig:dendrogram_perm} depicts the dendrograms for the same experiments shown in Figure~\ref{fig:dendrogram}, but with the vertical axes now corresponding to the time steps instead of the component size. The node labels are shown in the same order, for the sake of a more direct comparison between Figures~\ref{fig:dendrogram_perm} and~\ref{fig:dendrogram}. The length of each branch can now be understood to correspond to the \emph{permanence} of each respective component.

These results indicate that, as could be expected, larger connected components tend to last for longer periods of time than smaller components. While the dendrograms obtained for the ER and BA networks are mostly similar, the larger components of the GEO model present a tendency to last for a longer time.

\section{Concluding Remarks}

Complex networks, random walks, and hierarchy constitute three interesting research subjects that have been frequently addressed recently in the literature. In the present work, these three issues have been brought together relatively to the perspective of the gradual cutting of complex networks along respectively performed random walks, therefore establishing a respective hierarchy (binary tree) representable by a dendrogram. More specifically, given a network, one or more abstract agents perform a respective uniform random walk, and each traversed link is respectively removed from the network. Therefore, the networks are expected to undergo successive topological changes, sometimes breaking into exactly two connected components with respective sizes $m$ and $n$, up to the complete dismantling of the original network.

Two types of cuttings have been considered: sequential and parallel. These two types of dynamics may correspond to distinct practical situations, involving a single and multiple agents, respectively.

Among the several possible interesting issues implied by the considered slicing dynamics, we focused on two main specific questions, namely studying the balance and sizes of the components originating at each breakage (branching), as well as the permanence time of each of the involved connected components. 

The obtained results indicate that the considered specific type of GEO networks can have a substantially higher probability (than in the ER or BA cases) of yielding pairs of components that are not only more balanced (comparable sizes) but also relatively large. This result has several theoretical and practical situations regarding aspects including coverage and resilience of networks. For instance, interrupted parallel cutting dynamics may result in connected components whose size distribution could strongly depend on the type of network. Interestingly, all three types of networks have been found, at least for the adopted configurations, to have similar permanence times.

Several further works can be conceptualized in terms of the reported concepts, methods, and results. For instance, it would be interesting to investigate other types of random walks, including self-avoiding dynamics. Another interesting perspective would be to try to identify specific types of dynamics capable of yielding maximally or minimally balanced connected components. It would also be of interest to consider the presented concepts and methods respectively to other models of complex networks, including weighted, modular and/or real-world networks.

\section*{Acknowledgments}
Alexandre Benatti thanks MCTI PPI-SOFTEX (TIC 13 DOU 01245.010\\222/2022-44).
Luciano da F. Costa thanks CNPq (grant no.~307085/2018-0) and FAPESP (grant no.2022/15304-4) for financial support.

\bibliography{ref}
\bibliographystyle{unsrt}

\end{document}